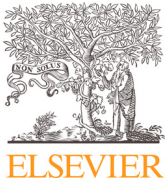
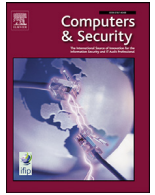

# DL-Droid: Deep learning based android malware detection using real devices

Mohammed K. Alzaylaee [a],[*], Suleiman Y. Yerima [b], Sakir Sezer [c]

[a] *College of Computing in Al-Qunfudah, Umm Al-Qura University, Saudi Arabia*
[b] *De Montfort University, Leicester, England, LE1 9BH, United Kingdom*
[c] *Centre for Secure Information Technologies (CSIT), Queen's University Belfast, Belfast BT7 1NN, United Kingdom*

**ABSTRACT**

The Android operating system has been the most popular for smartphones and tablets since 2012. This popularity has led to a rapid raise of Android malware in recent years. The sophistication of Android malware obfuscation and detection avoidance methods have significantly improved, making many traditional malware detection methods obsolete. In this paper, we propose DL-Droid, a deep learning system to detect malicious Android applications through dynamic analysis using stateful input generation. Experiments performed with over 30,000 applications (benign and malware) on real devices are presented. Furthermore, experiments were also conducted to compare the detection performance and code coverage of the stateful input generation method with the commonly used stateless approach using the deep learning system. Our study reveals that DL-Droid can achieve up to 97.8% detection rate (with dynamic features only) and 99.6% detection rate (with dynamic + static features) respectively which outperforms traditional machine learning techniques. Furthermore, the results highlight the significance of enhanced input generation for dynamic analysis as DL-Droid with the state-based input generation is shown to outperform the existing state-of-the-art approaches.

## 1. Introduction

Android operating system, which is provided by Google, is predicted to continue have a dramatic increase in the market with around 1.5 billion Android-based devices to be shipped by 2021 sta. It is currently leading the mobile OS market with over 80% market share compared to iOS, Windows, Blackberry, and Symbian OS. The availability of diverse Android markets such as Google Play, the official store, and third-party markets makes Android devices a popular target to not only legitimate developers, but also malware developers. Over one billion devices have been sold and more than 65 billion downloads have been made from Google Play (Smartphone, 0000). Android apps can be found in different categories, such as educational apps, gaming apps, social media apps, entertainment apps, banking apps, etc.

As a technology that is open source and widely adopted, Android is facing many challenges especially with malicious applications. The malware infected apps have the ability to send text messages to premium rate numbers without the user acknowledgment, gain access to private data, or even install code that can download and execute additional malware on the victim's device. The malware can also be used to create mobile botnets (Anagnostopoulos et al., 2016). Over the last few years, the number of malware samples attacking Android has significantly increased. According to a recent report from McAfee, over 2.5 million new Android malware apps were discovered in 2017, thus increasing the number of mobile malware samples in the wild to almost 25 million in 2017 (McA, 0000).

In order to mitigate the spread of malware, Google introduced a detection mechanism to its app market in Feb 2012 called Bouncer. Bouncer tests submitted applications in a sandbox for five minutes in order to detect any harmful behaviours; however, it has been shown that bouncer can be evaded by means of some simple detection avoidance methods (Oberheide and Miller, 2012). Alongside Bouncer, Google introduced Google Play protect in the Google 2017 event (Google Play, 2018). Google Play Protect is an always-on service that scans the applications automatically even after installation to ensure that the installed applications remains harmless 24/7. It has been reported that over 50 billion apps are scanned and verified every day regardless of where they were download from. However, according to McAfee, Google Play Protect also failed when tested against malware discovered in the previous 90 days in 2017 (McA, 0000). Furthermore, most third-party stores do not have the capability to scan and detect submitted harmful applications. Clearly, there is still a need for additional research into efficient methods to detect zero-day Android malware in the wild in order to overcome the aforementioned challenges.

* Corresponding author.
  *E-mail addresses:* mkzaylaee@uqu.edu.sa (M.K. Alzaylaee), syerima@dmu.ac.uk (S.Y. Yerima), s.sezer@qub.ac.uk (S. Sezer).

Various approaches have been proposed in previous works with the intention of detecting Android malware. These approaches are categorized into static analysis, dynamic analysis or hybrid analysis (where static and dynamic are used together). The methods based on static analysis reverse engineers the application for malicious code analysis. Arp et al. (2014), Aafer et al. (2013), Yerima et al. (2015a), Fan et al. (2017), Yerima et al. (2015b), Kang et al. (2016b), Cen et al. (2015), Westyarian et al. (2015) and Kang et al. (2016a) are few examples of detection methods using static analysis. By contrast, dynamic analysis executes the application in a controlled environment such as an emulator, or a real device with the purpose of tracing its behaviour. Several dynamic approaches, such as Enck et al. (2010); Alzaylaee, M.K., Yerima, S. Y., and Sezer S. (2016) DroidBox; Rastogi et al. (2013); Tam et al. (2015)tra, NVISO have been proposed. However, the efficiency of these approaches rely on the ability to detect the malicious behaviour during the runtime while providing the perfect environment to kick-start the malicious code.

Deep learning (DL) has gained increasing attention in the machine learning community and is re-emerging as a popular method of AI being applied in many fields (Hou et al., 2016; 2017; Karbab et al., 2017; LeCun et al., 2015; McLaughlin et al., 2017; Yuan et al., 2014; 2016). DL classifiers have inspired a great number of effective approaches in image classification, natural language processing, and speech recognition. Recently, Android malware researchers have also been exploring DL classifiers for malware analysis in order to increase detection accuracy.

Contrary to previous deep learning based dynamic detection works, this paper proposes and investigates a new system that exploits the advantages of deep learning coupled with dynamic stateful input generation, with the objective of achieving higher accuracy detection of zero-day Android malware. Furthermore, several experiments are conducted using real devices to compare the performance of the proposed DL based approach with those of popular machine learning classifiers. In summary, the main contributions of this paper are as follows:

- We present DL-Droid, a deep learning-based dynamic analysis system for Android malware detection. Unlike existing dynamic analysis systems, DL-Droid utilizes a state-based input generation approach for enhanced code coverage thus enabling improved performance.
- Using DL-Droid, we investigate the performance of the stateful input generation approach by utilizing the state-of-the-practice stateless (random-based) input generation as a comparative baseline. Higher accuracies were obtained with the stateful approach, thus highlighting the significance of enhanced input generation for Android malware detection systems that utilize dynamic analysis.
- We present an extensive comparative study of DL-Droid with seven popular machine learning classifiers. Unlike most existing studies that are based on emulators, our experiments are conducted in a more realistic environment using real devices. Experimental results show that DL-Droid outperforms the accuracy of traditional classifiers.

The rest of the paper is structured as follows. Section 2 discusses the related work. Followed by the methodology and experiments undertaken to evaluate DL-Droid in Section 3 Section 4 presents detailed experimental results and discussions of these results. Followed by the conclusion in Section 5

## 2. Related work

This section discusses the related work on Android malware detection, automated test input generation for Android, and recent works on deep learning approaches. As mentioned earlier, detecting Android malware with static analysis, where the application will be disassembled to be examined for presence of any malicious code is a popular approach. Several solutions have been developed using the static approach, utilizing features such as permissions, API calls, commands, and Intents. (Aafer et al., 2013; Arp et al., 2014; Cen et al., 2015; Fan et al., 2017; Yerima and Sezer, 2019; Yerima et al., 2016a; 2015a; 2015b), are examples of detection solutions based on static analysis. Although static analysis approaches enable more extensive code coverage, malware developers can use obfuscation techniques to hide the malicious code in order to evade static analysis. For example, data encryption, obfuscation, update attacks or polymorphic techniques. Therefore, in this work we only extract Android permissions statically prior to each run, and then extract the API calls and Intents dynamically at run time.

Dynamic analysis approach on the other hand, consist of running Android applications in a controlled environment such as an Android Virtual Device (AVD) emulator emu, or Genymotion Gen or in a real device in order to monitor the apps' behaviour. Alzaylaee et al. (2017) showed that analysing Android application in real phones is more effective in terms of stability and detecting more features compared to the emulator environment. Therefore, we chose to run our analysis on features extracted from real devices instead of emulators.

Automated dynamic analysis of Android apps requires streams of user emulated input events such as touches, gestures, or clicks to enable greater code coverage when run in either emulator or real phone. Choudhary et al. (2015) demonstrated that among the input generation tools analysed comparatively in their study (i.e. Monkey Developers (2012), Dynodroid Machiry et al. (2013), ACTEve Anand et al. (2012), A3E Azim and Neamtiu (2013), GUIRipper Amalfitano et al. (2012), SwiftHand Choi et al. (2013), and PUMA Hao et al. (2014)), Monkey performed the best in terms of code coverage. Nevertheless, in Alzaylaee et al. (2017) and Yerima et al. (2019) investigations proved that Monkey's code coverage capability could be surpassed by stateful approach enabled by tools such as DroidBot Li et al. (2017). The same studies have also shown that a stateful input generation is more stable and robust compared to the stateless approach enabled by Monkey. Hence, the deep learning-based system DL-Droid proposed in this paper is based on the dynamic stateful input generation approach.

The difficulty of detecting Android malware manually has led researches to explore the use of machine learning to automate and speed the detection process. Arp et al. (2014); Dini et al. (2012); Peiravian and Zhu (2013); Rasthofer et al. (2014); Shabtai et al. (2012); Yerima et al. (2015b, 2016b) are examples of published research that apply machine learning techniques to detect zero-day Android malware. Deep learning is re-emerging as a machine learning approach that is growing in popularity in many fields including Android malware detection. Droid-Sec Yuan et al. (2014) is one of the first frameworks that applied deep learning to classify Android malware, achieving 96.5% accuracy using 200 features extracted by means of a hybrid (static + dynamic) approach evaluated on 250 clean and 250 malware Android apps. Droid-Sec was a preliminary work for DroidDetector Yuan et al. (2016), where the authors increased the number of the analysed apps to 20,000 clean and 1760 malware and achieved 96.76% accuracy. Hou et al. (2016) proposed *Deep4MalDroid*, an automatic Android malware detection system, which will dynamically extract Linux kernel system calls using Genymotion emulator. The best detection accuracy they reached was 93.68% on features extracted from 1500 benign and 1500 malware Android apps. Similarly, Hou et al. (2017) proposed *AutoDroid* (automatic Android malware detection) based on API calls extracted using static analysis. Their system was developed using different types of deep neural networks (i.e., DBN and SAEs). The best accuracy of the DBN was 95.98% based on experiments with 2500 benign and 2500 malware Android apps.

In contrast to existing deep learning based Android malware detection frameworks, the key differentiates of our proposed DL-Droid framework is its dynamic stateful input generation approach. Furthermore, our work is based on real devices rather than emulators. Moreover, we employed 420 static and dynamic features and achieved better performance than existing frameworks. To the best of our knowledge, this is the first work that extensively investigates Android malware on real devices using over 30,000 applications, and presents evaluations with different input generation methods in order to measure the impact of their code coverage capabilities on the proposed DL-based malware detection approach.

Since our approach is based on feature extraction from real devices instead of emulators, the system is inherently robust against detection avoidance techniques aimed at emulators. Dynamic extraction from real devices also enables the system to overcome the limitations of static analysis e.g. dynamic code loading, obfuscation, data encryption, etc. It is also worth noting that some of the 420 features extracted are indicative of the malware incorporating these evasive behaviours, thus the deep learning system will be automatically equipped with the ability to learn how to detect malware with these behaviours during the training phase.

## 3. Methodology and experiments

In this section, we describe the methodology and the experiments which were conducted in order to evaluate the performance of the DL-Droid approach using real phones and two different test input generation methods: Stateless and Stateful.

### 3.1. Experimental setup

An automated platform is needed to run Android apps and extract their features. These features will be used as inputs for DL-Droid's deep learning based classification in order to detect Android malware. Since our aim is to investigate the performance of DL-Droid through several experiments, we utilized the DynaLog dynamic analysis framework described in Alzaylaee, M.K., Yerima, S.Y. and Sezer, S. (2016).

DynaLog is designed to accept and run a large number of Android apps automatically, launch them in sequence using either an emulator (an Android Virtual Device "AVD") or a real phone, and log and extract several dynamic features (i.e. API calls, Action/events). With the dynamic analysis of Android apps, test input generation is needed in order to ensure sufficient code coverage to trigger the malicious behaviours. DynaLog is capable of utilizing different test input generation methods including: stateless (random-based) (using the Monkey tool Developers (2012)), stateful (using DroidBot Li et al. (2017)), and hybrid-based (which combines stateless and stateful input generation tools Alzaylaee et al. (2017). The stateless approach is the most popular input generation approach and have been used extensively by researchers in this field. In fact, most existing dynamic analysis platforms for Android malware detection utilize a stateless approach (based on the Monkey tool). A previous study Yerima et al. (2019), compared the performance of stateless, stateful and hybrid-based input generation on various machine learning classifiers. In this study, the stateful approach is proved to be more robust and enabled greater code coverage than the hybrid-based input generation. Therefore, in this paper, we used only stateless (Monkey) and stateful (DroidBot) input generation for our experiments with DL-Droid.

The Monkey tool generates pseudo-random streams of events because of a pseudo-random number generator which is controlled by a seed. A pseudo-random stream of events is still a random approach since the event selection is not based on a pre-determined pattern, even though it is configurable. It is important to note that random here refers to selection of next event to be executed i.e. no specific pattern is followed, unlike in the stateful approach where the event to be executed is chosen based on evaluation of the current state (i.e. the user interfaces state at a particular time). Monkey is based on a stateless approach and this is the most important difference that distinguishes the approach from the stateful DroidBot. We propose the stateful approach the default component of the DL-Droid framework, while we utilize the stateless method as a baseline for comparative analysis in this paper.

Alzaylaee et al. (2017) have shown that dynamic analysis done on real devices is more efficient than using emulators. Thus, our experiments are completely based on real phones. Eight different phone brands were used with the following configurations: Android 6.0 "Marshmallow", 4GB RAM, 2.6Hz CPU, 32GB ROM and 32 GB of external SD card storage. Each smartphone processed an average of 100 apps daily. The SD cards contained a folder full of different resources such as pictures, text files, videos, sound files, etc. to simulate a typical phone. Moreover, each phone was equipped with a sim card containing call credits to enable 3G data usage, send text messages, and even make a phone calls when requested. The phones were also connected to an internal WiFi service in order to enable tested applications to connect with their external servers when necessary.

The executed runtime was different on each run and determined by the chosen test input generation. The required timing was confirmed after evaluation with several apps to determine how much time was needed to trigger every possible event using either the stateful tool (DroidBot), or the stateless tool (Monkey). For the stateful method, 180 s was found to be sufficient. For the stateless generation using Monkey, 300 s was enough to generate 4000 events for the apps. Beyond 4000 events, most apps did not generate any further dynamic output from Dynalog. The overview of the DL-Droid process using DynaLog as well as the DL classifier engine is shown in Fig. 1

### 3.2. Dataset

For the purpose of evaluating DL-Droid accuracy performance and to compare it with other popular machine learning classifiers, we used a dataset consisting 31,125 Android applications. Out of these, 11,505 were malware samples while the rest were 19,620 internally vetted benign samples obtained from Intel Security (McAfee Labs). These samples consist of a variety of app formats, including paid apps, powerful utility apps, banking apps, media player apps, and popular game apps. The samples are available to other researchers on request.

### 3.3. Features extraction and preprocessing

In the feature extraction phase, each application is installed and run on one of the eight phones using DynaLog (Alzaylaee, M.K., Yerima, S. Y. and Sezer, S. 2016). Once completed for each of the two scenarios, Stateless and Stateful, the logged features are preprocessed into text files of feature vectors representing the features extracted from each application. These text files were further processed into a single.csv file for each scenario with the purpose of evaluating the detection performance using deep learning. The.csv is an acceptable file format for both H2O flow and WEKA which were used later in the experiments. Note that, each feature in the.csv file is binary containing either '0' or '1' which represents the absence or presence of each extracted features.

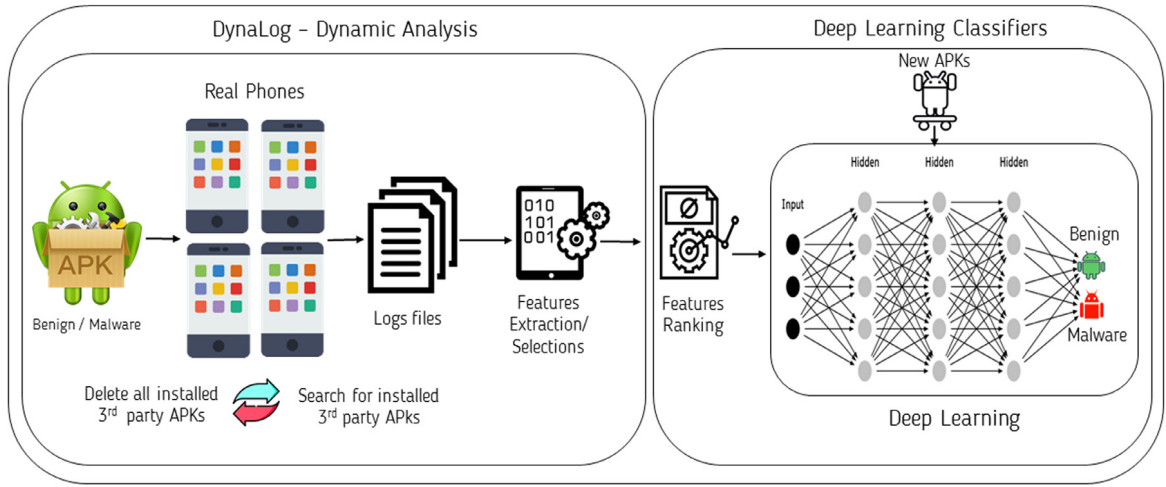

**Fig. 1.** DL-Droid framework.

**Table 1**
Total number of the extracted features used in the experiments.

| Feature set | No. of features |
|---|---|
| Application attributes features | 97 |
| Actions/Events features | 23 |
| Permission features | 300 |
| **Total** | **420** |

Originally, DynaLog was equipped to extract 178 features dynamically (i.e. API calls and intents - Actions/Events). These features were ranked using InfoGain (information gain provided by WEKA), and then the top 120 features were selected for the experiments. Dynalog was extended to enable extracting Android permissions statically prior to each dynamic run. This step allows us to test the detection performance of DL-Droid with more features as a bonus.

Over 300 Android permissions that have been used by the investigated Android apps that govern access to different device hardware and system resources. These permissions are considered as either normal, signature, or dangerous permissions. This step allowed us to collect the most relevant permissions which some of which were relatively new and had not been used in previous works. Hence, as shown in Table 1, we obtained a total of 420 features from our feature extraction phase. The 420 extracted features were ranked using the information gain (InfoGain) feature ranking algorithm in WEKA. The top 20 ranked dynamic features (including and excluding the extracted permissions) based on InfoGain in both test input scenarios (stateless and stateful) are shown in Tables 2–5 respectively.

### 3.4. Features ranking comparisons

From Tables 2, and–5, it is interesting to note that the API calls methods getDeviceId, getSubscriberId, getLine1Number, and getSimSerialNumber from the TelephonyManager class, that provides access to information about the telephony services on the device, were among the top 20. However, the InfoGain score is higher for these features when extracted using DroidBot-based stateful input generation. For example, the InfoGain score of the feature TelephonyManager;-> getDeviceId is 0.099 in Tables 2 and 3 using DroidBot based stateful input generation, whereas the score for the same feature using stateless Monkey based random input generation is 0.075 in Tables 4 and 5.

Similar findings can be seen with the feature TelephonyManager;-> getSubscriberId which scored 0.057 using DroidBot based input generation, while the score is 0.042 for Monkey based input generation. The feature action.SMS_RECEIVED scores 0.096 for the DroidBot based generation in Table 2, which is higher than the score for the same feature extracted using Monkey based generation in Table 4 Hence, this indicates that the stateful DroidBot based input generation method has triggered more behaviours than the stateless Monkey based random input generation. Note that most existing dynamic analysis on Android utilize the Monkey tool for input event generation.

### 3.5. Investigating Deep Learning Classifier vs. other popular machine learning algorithms

Our main goal is to build a model for DL-Droid to enable accurate classification and detection of Android malware from benign apps. In our experiments, we train our deep learning classifiers on a classification problem with two labels, benign or malicious. We utilize H2O which currently supports only the Multilayer Perceptron classifier (MLP) Candel et al. (2016). A confusion matrix is performed in our system to evaluate the effectiveness of different classifiers. The second phase of the experiments compared the performance between the proposed DL and seven popular machine learning approaches proposed in the literature. The classifiers include: Support Vector Machine (SVM Linear), Support Vector Machine with radial basis function kernel (SVM RBF), Naive Bayes (NB), Simple Logistic (SL), Partial Decision Trees (PART), Random Forest (RF), and J48 Decision Tree. We also investigated the performance of each classifier for two different test input generation methods. The results of our experiments are presented in section III using the performance metrics defined as follows:

The true positive ratio (TPR) also known as recall, true negative ratio (TNR), false positive ratio (FPR), false negative ratio (FNR), and precision are defined as follows:

$$TPR = \frac{TP}{TP+FN} \quad (1)$$

$$TNR = \frac{TN}{TN+FP} \quad (2)$$

$$FPR = \frac{FP}{FP+TN} \quad (3)$$

**Table 2**

Top 20 Ranked Features based on InfoGain using Stateful input generation DroidBot (Permissions excluded).

| Feature | Malware | Bening | InfoGain score |
| --- | --- | --- | --- |
| TelephonyManager;->getDeviceId | 4899 | 1078 | 0.099 |
| com.android.vending.INSTALL_REFERRER | 741 | 3890 | 0.096 |
| action.SMS_RECEIVED | 2421 | 375 | 0.096 |
| TelephonyManager;->getSubscriberId | 1993 | 203 | 0.057 |
| action.USER_PRESENT | 2633 | 545 | 0.056 |
| methods/HttpPost;-><init> | 3408 | 1044 | 0.054 |
| TelephonyManager;->getLine1Number | 1429 | 163 | 0.043 |
| WifiManager;->getConnectionInfo | 2680 | 792 | 0.039 |
| content/Context;->bindService | 573 | 2271 | 0.037 |
| Ljava/util/TimerTask;-><init> | 7399 | 4068 | 0.033 |
| Ljava/io/FileOutputStream;->write | 3775 | 1563 | 0.032 |
| PackageManager;->checkPermission | 2726 | 858 | 0.032 |
| Landroid/net/NetworkInfo;->getState | 1632 | 396 | 0.032 |
| Ljava/io/File;->exists | 6217 | 3361 | 0.027 |
| security/MessageDigest;->getInstance | 4905 | 2779 | 0.027 |
| Landroid/content/Context;->unbindService | 264 | 1347 | 0.02 |
| action.PHONE_STATE | 1030 | 215 | 0.019 |
| action.PACKAGE_ADDED | 1540 | 508 | 0.018 |
| TelephonyManager;->getSimSerialNumber | 859 | 157 | 0.018 |
| SmsManager;->sendTextMessage | 351 | 2 | 0.018 |

**Table 3**

Top 20 Ranked Features based on InfoGain using Stateful input generation DroidBot (Permissions included).

| Feature | Malware | Bening | InfoGain score |
| --- | --- | --- | --- |
| permission.SEND_SMS | 5128 | 1084 | 0.16 |
| permission.READ_PHONE_STATE | 10,508 | 10,183 | 0.133 |
| TelephonyManager;->getDeviceId | 4899 | 2011 | 0.099 |
| com.android.vending.INSTALL_REFERRER | 741 | 7285 | 0.096 |
| permission.RECEIVE_SMS | 4054 | 1565 | 0.096 |
| action.MOUNT_UNMOUNT_FILESYSTEMS | 2889 | 781 | 0.082 |
| permission.WRITE_SMS | 2847 | 764 | 0.071 |
| permission.READ_SMS | 3592 | 1429 | 0.07 |
| permission.SYSTEM_ALERT_WINDOW | 4314 | 2276 | 0.069 |
| action.SMS_RECEIVED | 2421 | 665 | 0.066 |
| TelephonyManager;->getSubscriberId | 1993 | 387 | 0.057 |
| action.USER_PRESENT | 2633 | 912 | 0.056 |
| permission.INSTALL_PACKAGES | 1640 | 290 | 0.054 |
| permission.ACCESS_MTK_MMHW | 1092 | 29 | 0.047 |
| permission.GET_TASKS | 5790 | 5040 | 0.046 |
| permission.RECEIVE_BOOT_COMPLETED | 6648 | 6378 | 0.044 |
| methods/HttpPost;-><init> | 3408 | 1985 | 0.044 |
| permission.USE_CREDENTIALS | 313 | 3369 | 0.043 |
| TelephonyManager;->getLine1Number | 1429 | 278 | 0.041 |
| permission.ACCESS_WIFI_STATE | 8406 | 9802 | 0.039 |

**Table 4**

Top 20 Ranked Features based on InfoGain using stateless Monkey based input generation (Permissions excluded).

| Feature | Malware | Bening | InfoGain score |
| --- | --- | --- | --- |
| com.android.vending.INSTALL_REFERRER | 737 | 6686 | 0.106 |
| TelephonyManager;->getDeviceId | 4269 | 1800 | 0.075 |
| action.SMS_RECEIVED | 2416 | 608 | 0.057 |
| action.USER_PRESENT | 2627 | 853 | 0.053 |
| TelephonyManager;->getSubscriberId | 1682 | 363 | 0.042 |
| TelephonyManager;->getLine1Number | 1385 | 286 | 0.035 |
| Landroid/content/Context;->bindService | 589 | 3260 | 0.031 |
| Landroid/net/NetworkInfo;->getState | 1570 | 593 | 0.026 |
| client/methods/HttpPost;-><init> | 2526 | 1634 | 0.022 |
| Ljava/io/FileOutputStream;->write | 3082 | 2296 | 0.021 |
| Ljava/util/TimerTask;-><init> | 6943 | 7564 | 0.021 |
| action.PHONE_STATE | 1026 | 349 | 0.018 |
| TelephonyManager;->getSimSerialNumber | 896 | 257 | 0.018 |
| Landroid/content/Context;->unbindService | 287 | 1800 | 0.018 |
| wifi/WifiManager;->getConnectionInfo | 1977 | 1235 | 0.018 |
| action.PACKAGE_ADDED | 1533 | 804 | 0.017 |
| /PackageManager;->checkPermission | 2111 | 1419 | 0.017 |
| Ljava/io/File;->exists | 5674 | 6027 | 0.016 |
| telephony/SmsManager;->sendTextMessage | 332 | 1 | 0.015 |
| action.NEW_OUTGOING_CALL | 655 | 182 | 0.013 |

**Table 5**
Top 20 Ranked Features based on InfoGain using stateless Monkey based input generation (Permissions included).

| Feature | Malware | Bening | InfoGain score |
|---|---|---|---|
| permission.SEND_SMS | 5117 | 1006 | 0.16 |
| permission.READ_PHONE_STATE | 10,468 | 9146 | 0.135 |
| android.vending.INSTALL_REFERRER | 737 | 6686 | 0.106 |
| permission.RECEIVE_SMS | 4043 | 1420 | 0.082 |
| TelephonyManager;-> getDeviceId | 4269 | 1800 | 0.075 |
| permission.WRITE_SMS | 2841 | 697 | 0.07 |
| action.MOUNT_UNMOUNT_FILESYSTEMS | 2879 | 741 | 0.069 |
| permission.READ_SMS | 3584 | 1322 | 0.068 |
| permission.SYSTEM_ALERT_WINDOW | 4299 | 2138 | 0.063 |
| action.SMS_RECEIVED | 2416 | 608 | 0.057 |
| action.USER_PRESENT | 2627 | 853 | 0.053 |
| permission.INSTALL_PACKAGES | 1638 | 243 | 0.049 |
| permission.ACCESS_MTK_MMHW | 1087 | 26 | 0.046 |
| permission.GET_TASKS | 5770 | 4534 | 0.044 |
| permission.RECEIVE_BOOT_COMPLETED | 6621 | 5760 | 0.043 |
| permission.USE_CREDENTIALS | 311 | 3004 | 0.043 |
| TelephonyManager;-> getSubscriberId | 1682 | 363 | 0.042 |
| permission.GET_ACCOUNTS | 3012 | 8601 | 0.039 |
| permission.ACCESS_WIFI_STATE | 8376 | 8941 | 0.035 |
| TelephonyManager;-> getLine1Number | 1385 | 286 | 0.035 |

$$FNR = \frac{FN}{FN + TP} \quad (4)$$

$$P = \frac{TP}{TP + FP} \quad (5)$$

Where TP denotes the number of true positives, TN the number of true negatives, FP the number of false positives, and FN the number of false negatives.

FM is the F measure calculated for both malware and benign classes. The combined measure known as weighted FM is defined as follows:

$$FM = \frac{2 * recall * precision}{recall + precision} \quad (6)$$

$$W - FM = \frac{(F_m.N_m) + (F_b.N_b)}{N_m + N_b} \quad (7)$$

Where $F_b$ and $F_m$ are the FM of the benign and malware datasets respectively, whereas $N_b$ and $N_m$ are the number of samples in the benign and malware datasets respectively. The 10-fold cross validation approach was used in all of the presented experiments.

## 4. Experimental results and discussions

### 4.1. Deep learning classifier analysis

#### 4.1.1. DL comparisons with dynamic features: Stateful vs. Stateless input generation

Table 6 depicts the results of experiments undertaken to evaluate the performance of the DL approach with different combinations of hidden layers. The results shown here is for the dynamic features only, using the stateful Droidbot input generation tool. 22 different combinations of hidden neurons, containing two, three, and four layers, have been applied in order to determine the best possible performance based on the w-FM. At Table 6, the results show that the 200, 200, 200 combination performs the best when compared to other combinations, with running time of nine min-utes. We can see that DL can achieve w-FM of 0.963 when setting the number of layers to 3 and selecting 200 neurons in each layer with dynamic features only.

**Table 6**
Deep learning results with different combinations of hidden layers (with the use of stateful input generation and dynamic features only).

| No. of layers | No. of Neurons | TPR | TNR | FPR | FNR | Precision | Recall | Accuracy | w-FM | AUC | Running time (min:sec) |
|---|---|---|---|---|---|---|---|---|---|---|---|
| 2 | 50,50 | 0.9532 | 0.8804 | 0.1196 | 0.0468 | 0.9317 | 0.9532 | 0.9264 | 0.9423 | 0.971476 | 01:35 |
| 2 | 100,100 | 0.9337 | 0.8896 | 0.1104 | 0.0663 | 0.9374 | 0.9337 | 0.9178 | 0.9355 | 0.965159 | 03:14 |
| 2 | 200,200 | 0.9663 | 0.9075 | 0.0925 | 0.0337 | 0.9479 | 0.9663 | 0.9449 | 0.957 | 0.981357 | 06:31 |
| 2 | 300,300 | 0.9663 | 0.907 | 0.093 | 0.0337 | 0.9469 | 0.9663 | 0.9445 | 0.9565 | 0.982852 | 10:46 |
| 2 | 400,400 | 0.9903 | 03:04.3 | 0.2062 | 0.0097 | 0.8889 | 0.9903 | 0.9166 | 0.9369 | 0.895044 | 13:44 |
| 2 | 500,500 | 0.9739 | 0.8975 | 0.1025 | 0.0261 | 0.9428 | 0.9739 | 0.946 | 0.9581 | 0.982761 | 18:46 |
| 3 | 50,50,50 | 0.9047 | 0.9082 | 0.0918 | 0.0953 | 0.9447 | 0.9047 | 0.906 | 0.9243 | 0.928607 | 01:50 |
| 3 | 100,50,100 | 0.973 | 0.8929 | 0.1071 | 0.027 | 0.94 | 0.973 | 0.9436 | 0.9562 | 0.982075 | 03:19 |
| 3 | 50,100,50 | 0.9762 | 0.8832 | 0.1168 | 0.0238 | 0.934 | 0.9762 | 0.9417 | 0.9546 | 0.979091 | 02:14 |
| 3 | 100,100,100 | 0.7478 | 0.9542 | 0.0458 | 0.2522 | 0.9649 | 0.7478 | 0.8246 | 0.8426 | 0.861435 | 03:50 |
| 3 | 100,200,100 | 0.5356 | 0.9868 | 0.0132 | 0.4644 | 0.9854 | 0.5356 | 0.7046 | 0.694 | 0.76359 | 05:06 |
| 3 | 200,100,200 | 0.4526 | 0.9997 | 0.0003 | 0.5474 | 0.9996 | 0.4526 | 0.6536 | 0.6231 | 0.726156 | 06:42 |
| **3** | **200,200,200** | **0.9776** | **0.9086** | **0.0914** | **0.0224** | **0.9482** | **0.9776** | **0.9521** | **0.9627** | **0.986742** | **09:05** |
| 3 | 300,100,300 | 0.5384 | 0.9986 | 0.0014 | 0.4616 | 0.9985 | 0.5384 | 0.7086 | 0.6996 | 0.768651 | 09:08 |
| 3 | 300,300,300 | 0.9735 | 0.9055 | 0.0945 | 0.0265 | 0.9445 | 0.9735 | 0.9479 | 0.9588 | 0.9838 | 13:56 |
| 3 | 400,400,400 | 0.7219 | 0.9842 | 0.0158 | 0.2781 | 0.9875 | 0.7219 | 0.8183 | 0.8341 | 0.85698 | 20:40 |
| 3 | 500,500,500 | 0.9738 | 0.9089 | 0.0911 | 0.0262 | 0.9478 | 0.9738 | 0.9497 | 0.9606 | 0.984606 | 26:44 |
| 4 | 50,50,50,50 | 0.9567 | 0.8949 | 0.1051 | 0.0433 | 0.9397 | 0.9567 | 0.9339 | 0.9482 | 0.97666 | 01:58 |
| 4 | 100,100,100,100 | 0.5671 | 0.9986 | 0.0014 | 0.4329 | 0.9986 | 0.5671 | 0.7248 | 0.7234 | 0.783015 | 04:24 |
| 4 | 200,200,200,200 | 0.9744 | 0.9109 | 0.0891 | 0.0256 | 0.9488 | 0.9744 | 0.9508 | 0.9614 | 0.984959 | 10:39 |
| 4 | 300,300,300,300 | 0.9868 | 0.8552 | 0.1448 | 0.0132 | 0.9206 | 0.9868 | 0.9381 | 0.9525 | 0.921231 | 18:39 |
| 4 | 400,400,400,400 | 0.9622 | 0.885 | 0.115 | 0.0378 | 0.9353 | 0.9622 | 0.9339 | 0.9485 | 0.975693 | 29:28 |

**Table 7**
Deep learning results with different combinations of hidden layers (with the use of stateful input generation and static + dynamic features).

| No. of layers | No. of Neurons | TPR | TNR | FPR | FNR | Precision | Recall | Accuracy | w-FM | AUC | Running time (min:sec) |
|---|---|---|---|---|---|---|---|---|---|---|---|
| 2 | 50,50 | 0.8701 | 0.9252 | 0.0748 | 0.1299 | 0.952 | 0.8701 | 0.8904 | 0.9092 | 0.920539 | 03:43 |
| 2 | 100,100 | 0.9669 | 0.9162 | 0.0838 | 0.0331 | 0.9516 | 0.9669 | 0.9482 | 0.9592 | 0.980748 | 07:18 |
| 2 | 200,200 | 0.9661 | 0.8771 | 0.1229 | 0.0339 | 0.9306 | 0.9661 | 0.9332 | 0.948 | 0.970208 | 14:51 |
| 2 | 300,300 | 0.9693 | 0.9121 | 0.0879 | 0.0307 | 0.9495 | 0.9693 | 0.9481 | 0.9593 | 0.980417 | 21:42 |
| 2 | 400,400 | 0.972 | 0.9082 | 0.0918 | 0.028 | 0.9475 | 0.972 | 0.9484 | 0.9596 | 0.978372 | 30:50 |
| 2 | 500,500 | 0.9735 | 0.9261 | 0.0739 | 0.0265 | 0.9574 | 0.9735 | 0.956 | 0.9654 | 0.983072 | 36:36 |
| 3 | 50,50,50 | 0.9718 | 0.9059 | 0.0941 | 0.0282 | 0.9462 | 0.9718 | 0.9474 | 0.9588 | 0.978387 | 04:05 |
| 3 | 100,50,100 | 0.9718 | 0.9149 | 0.0851 | 0.0282 | 0.9512 | 0.9718 | 0.9507 | 0.9613 | 0.983224 | 07:38 |
| 3 | 50,100,50 | 0.9646 | 0.8707 | 0.1293 | 0.0354 | 0.9271 | 0.9646 | 0.9299 | 0.9455 | 0.976274 | 04:34 |
| 3 | 100,100,100 | 0.9733 | 0.9202 | 0.0798 | 0.0267 | 0.9541 | 0.9733 | 0.9537 | 0.9636 | 0.982417 | 08:33 |
| 3 | 100,200,100 | 0.9737 | 0.9231 | 0.0769 | 0.0263 | 0.9557 | 0.9737 | 0.955 | 0.9647 | 0.984257 | 09:40 |
| 3 | 200,100,200 | 0.9733 | 0.9314 | 0.0686 | 0.0267 | 0.9603 | 0.9733 | 0.9578 | 0.9668 | 0.984114 | 15:13 |
| **3** | **200,200,200** | **0.9956** | **0.967** | **0.033** | **0.0044** | **0.9809** | **0.9956** | **0.985** | **0.9882** | **0.997105** | **17:21** |
| 3 | 300,100,300 | 0.976 | 0.9089 | 0.0911 | 0.024 | 0.9481 | 0.976 | 0.9512 | 0.9618 | 0.980004 | 22:34 |
| 3 | 300,300,300 | 0.9762 | 0.912 | 0.088 | 0.0238 | 0.9498 | 0.9762 | 0.9525 | 0.9628 | 0.982277 | 26:29 |
| 3 | 400,400,400 | 0.9764 | 0.9166 | 0.0834 | 0.0236 | 0.9523 | 0.9764 | 0.9543 | 0.9642 | 0.982026 | 38:16 |
| 3 | 500,500,500 | 0.9765 | 0.9287 | 0.0713 | 0.0235 | 0.959 | 0.9765 | 0.9588 | 0.9676 | 0.983941 | 49:31 |
| 4 | 50,50,50,50 | 0.9676 | 0.9195 | 0.0805 | 0.0324 | 0.9535 | 0.9676 | 0.9498 | 0.9605 | 0.981414 | 04:31 |
| 4 | 100,100,100,100 | 0.9661 | 0.9265 | 0.0735 | 0.0339 | 0.9573 | 0.9661 | 0.9515 | 0.9617 | 0.982657 | 09:29 |
| 4 | 200,200,200,200 | 0.9757 | 0.9207 | 0.0793 | 0.0243 | 0.9545 | 0.9757 | 0.9553 | 0.965 | 0.982594 | 20:05 |
| 4 | 300,300,300,300 | 0.9739 | 0.9131 | 0.0869 | 0.0261 | 0.9503 | 0.9739 | 0.9514 | 0.9619 | 0.980652 | 36:36 |
| 4 | 400,400,400,400 | 0.9717 | 0.9093 | 0.0907 | 0.0283 | 0.9481 | 0.9717 | 0.9486 | 0.9597 | 0.982227 | 43:52 |

We repeated the same experiments on the dynamic features extracted using the stateless Monkey based random input generation tool in order to compare the results with the previous scenario. Table 8 shows the results obtained. The best w-FM is also recorded with three layers similar to the previous scenario. However, this is obtained with different combination of neurons. The number of neurons in each layer is 300, 100, 300 respectively for the best w-FM of 0.958. Even though the running time is 8 minutes, which is less by almost one minute, our focus has been the detection accuracy. Therefore, from Tables 6 and 8, we can confirm that the DL-Droid achieves its best performance with the features obtained from the use of the stateful input generation approach.

### 4.1.2. DL comparisons with dynamic features and static features: Stateless vs. Stateful input generation

The same experiments outlined in the previous section were repeated with the addition of static features, i.e. permissions, and results are shown in Table 7 We can see that the same combination of 200 neurons in each hidden layer with three hidden layers is superior to the other deep networks for Android malware detection using the stateful input generation approach. The w-FM reached approximately 0.99.

### 4.2. Comparison of the performance of the Deep Learning Classifier with other popular machine learning classifiers

In this section, we compare the detection accuracies of the proposed DL approach with the most popular machine learning algorithms as shown in Tables 10 and 11. Overall seven machine learning algorithms were selected based on results of several preliminary experiments had been conducted. From the tables, we can clearly see that the proposed DL approach outperforms the other machine learning algorithms. In Table 10, where results from only dynamic features are presented, the second highest w-FM of 0.94 is achieved by the Random Forest algorithm, while that of the deep learning approach is 0.963. When we perform further comparison by adding permissions to the analysis (Table 11), the DL approach still topped the rest with a w-FM of nearly 0.99, while the next highest, which is again Random Forest, achieved a w-FM of 0.97. We can clearly observe that the addition of static features i.e. permissions improved the detection accuracy of DL-Droid.

Fig. 2, presents the results of comparison between the two input generation methods i.e. stateful (using Droidbot) and stateless (using monkey). Fig. 2 shows the w-FM results for DL-Droid as well as the selected seven popular machine learning algorithms. In the experiments with dynamic features only, all classifiers except for NB and J48, performed better where stateful input generation with Droidbot was utilized, compared to the stateless approach using Monkey. However, in the experiment with combined static and dynamic features, the stateful input generation approach was superior for all the classifiers. With these results depicted in Fig. 2, we can conclude that DL-Droid with stateful input generation (our initially proposed approach) achieves the best detection accuracy.

### 4.3. Results comparison with existing work

Table 12, presents a comparison of DL-Droid performance with other existing deep learning based methods for Android malware detection. DroidDetector's static and dynamic based deep learning method achieved 96.76% accuracy compared to DL-Droid which has 98.5% accuracy. DL-Droid outperformed DroidDetector (Yuan et al., 2016) in all other metrics, while utilizing more samples for the experiments. DL-Droid also outperforms Maldozer (Karbab et al., 2017), Deep4MalDroid (Hou et al., 2016), AutoDroid (Hou et al., 2017) and the CNN approach presented in (McLaughlin et al., 2017). It is interesting to note that, just like in Deep4MalDroid and AutoDroid, the number of the optimum hidden layers for DL-Droid is three.

## 5. Conclusion

In this paper, we presented DL-Droid, an automated dynamic analysis framework for Android malware detection. DL-Droid employs deep learning with a state-based input generation approach as the default method, although it has the capability to employ the state-of-the-practice popular Monkey tool (stateless method). We evaluated DL-Droid using 31,125 Android applications, 420 static and dynamic features, comparing its performance to traditional machine learning classifiers as well as existing DL-based frameworks. The presented results clearly demonstrate that DL-Droid

**Table 8**
Deep learning results with different combinations of hidden layers (with the use of stateless input generation and dynamic features only).

| No. of layers | No. of Neurons | TPR | TNR | FPR | FNR | Precision | Recall | Accuracy | w-FM | AUC | Running time (min:sec) |
|---|---|---|---|---|---|---|---|---|---|---|---|
| 3 | 100,100,100 | 0.6111 | 0.9954 | 0.0046 | 0.3889 | 0.9951 | 0.6111 | 0.7619 | 0.7572 | 0.803875 | 03:31 |
| 3 | 50,100,50 | 0.6285 | 0.967 | 0.033 | 0.3715 | 0.9659 | 0.6285 | 0.7648 | 0.7615 | 0.806743 | 02:00 |
| 2 | 400,400 | 0.726 | 0.9615 | 0.0385 | 0.274 | 0.9666 | 0.726 | 0.8189 | 0.8292 | 0.854101 | 12:33 |
| 2 | 100,100 | 0.8221 | 0.9105 | 0.0895 | 0.1779 | 0.9321 | 0.8221 | 0.8575 | 0.8736 | 0.886959 | 02:53 |
| 2 | 50,50 | 0.8795 | 0.8266 | 0.1734 | 0.1205 | 0.8867 | 0.8795 | 0.8587 | 0.8831 | 0.89001 | 01:26 |
| 3 | 500,500,500 | 0.997 | 0.7606 | 0.2394 | 0.003 | 0.8622 | 0.997 | 0.9025 | 0.9247 | 0.879508 | 06:10 |
| 3 | 100,50,100 | 0.9233 | 0.9081 | 0.0919 | 0.0767 | 0.9372 | 0.9233 | 0.9172 | 0.9302 | 0.940436 | 02:52 |
| 3 | 50,50,50 | 0.9526 | 0.8824 | 0.1176 | 0.0474 | 0.9257 | 0.9526 | 0.9249 | 0.9389 | 0.973505 | 01:36 |
| 4 | 300,300,300,300 | 0.9672 | 0.8975 | 0.1025 | 0.0328 | 0.9349 | 0.9672 | 0.9396 | 0.9508 | 0.981281 | 17:00 |
| 4 | 50,50,50,50 | 0.9671 | 0.8977 | 0.1023 | 0.0329 | 0.9355 | 0.9671 | 0.9397 | 0.951 | 0.979724 | 01:49 |
| 3 | 200,200,200 | 0.9764 | 0.8905 | 0.1095 | 0.0236 | 0.9302 | 0.9764 | 0.942 | 0.9527 | 0.984125 | 07:12 |
| 4 | 100,100,100,100 | 0.9752 | 0.8902 | 0.1098 | 0.0248 | 0.9311 | 0.9752 | 0.9415 | 0.9527 | 0.981828 | 04:14 |
| 2 | 500,500 | 0.971 | 0.901 | 0.099 | 0.029 | 0.9363 | 0.971 | 0.943 | 0.9533 | 0.982513 | 16:08 |
| 4 | 200,200,200,200 | 0.9791 | 0.8831 | 0.1169 | 0.0209 | 0.9293 | 0.9791 | 0.9417 | 0.9535 | 0.981191 | 09:53 |
| 4 | 400,400,400,400 | 0.973 | 0.8978 | 0.1022 | 0.027 | 0.9359 | 0.973 | 0.9433 | 0.9541 | 0.982963 | 25:48 |
| 3 | 300,300,300 | 0.9754 | 0.8945 | 0.1055 | 0.0246 | 0.9342 | 0.9754 | 0.9435 | 0.9543 | 0.980887 | 11:57 |
| 3 | 200,100,200 | 0.9717 | 0.903 | 0.097 | 0.0283 | 0.9384 | 0.9717 | 0.9444 | 0.9547 | 0.983906 | 05:50 |
| 2 | 200,200 | 0.9696 | 0.9077 | 0.0923 | 0.0304 | 0.9415 | 0.9696 | 0.9452 | 0.9553 | 0.982541 | 06:19 |
| 3 | 400,400,400 | 0.9759 | 0.8963 | 0.1037 | 0.0241 | 0.9355 | 0.9759 | 0.9446 | 0.9553 | 0.983396 | 16:45 |
| 3 | 100,200,100 | 0.9741 | 0.9104 | 0.0896 | 0.0259 | 0.943 | 0.9741 | 0.9488 | 0.9583 | 0.985178 | 04:23 |
| 2 | 300,300 | 0.9798 | 0.8997 | 0.1003 | 0.0202 | 0.9386 | 0.9798 | 0.9486 | 0.9588 | 0.984684 | 09:37 |
| **3** | **300,100,300** | **0.9778** | **0.9064** | **0.0936** | **0.0222** | **0.9408** | **0.9778** | **0.9495** | **0.9589** | **0.985202** | **08:08** |

**Table 9**
Deep learning results with different combinations of hidden layers (with the use of stateless input generation and static + dynamic features).

| No. of layers | No. of Neurons | TPR | TNR | FPR | FNR | Precision | Recall | Accuracy | w-FM | AUC | Running time (min:sec) |
|---|---|---|---|---|---|---|---|---|---|---|---|
| 2 | 50,50 | 0.8788 | 0.8905 | 0.1095 | 0.1212 | 0.9244 | 0.8788 | 0.8834 | 0.901 | 0.916957 | 03:32 |
| 2 | 200,200 | 0.9246 | 0.8612 | 0.1388 | 0.0754 | 0.9102 | 0.9246 | 0.8995 | 0.9174 | 0.956576 | 14:11 |
| 3 | 50,50,50 | 0.9443 | 0.8505 | 0.1495 | 0.0557 | 0.9058 | 0.9443 | 0.9071 | 0.9247 | 0.961333 | 03:46 |
| 3 | 50,100,50 | 0.9403 | 0.87 | 0.13 | 0.0597 | 0.9168 | 0.9403 | 0.9125 | 0.9284 | 0.963836 | 04:18 |
| 2 | 100,100 | 0.9569 | 0.8841 | 0.1159 | 0.0431 | 0.9263 | 0.9569 | 0.928 | 0.9413 | 0.971361 | 07:18 |
| 4 | 400,400,400,400 | 0.9703 | 0.8882 | 0.1118 | 0.0297 | 0.9296 | 0.9703 | 0.9377 | 0.9495 | 0.975414 | 44:27 |
| 2 | 300,300 | 0.9625 | 0.9063 | 0.0937 | 0.0375 | 0.9399 | 0.9625 | 0.9402 | 0.9511 | 0.978022 | 21:32 |
| 3 | 200,200,200 | 0.9609 | 0.9131 | 0.0869 | 0.0391 | 0.9439 | 0.9609 | 0.942 | 0.9524 | 0.977116 | 16:27 |
| 4 | 50,50,50,50 | 0.9653 | 0.906 | 0.094 | 0.0347 | 0.9399 | 0.9653 | 0.9418 | 0.9525 | 0.977894 | 04:12 |
| 3 | 300,100,300 | 0.9764 | 0.8921 | 0.1079 | 0.0236 | 0.9323 | 0.9764 | 0.943 | 0.9539 | 0.977185 | 21:11 |
| 4 | 100,100,100,100 | 0.9649 | 0.9125 | 0.0875 | 0.0351 | 0.9438 | 0.9649 | 0.9441 | 0.9542 | 0.979354 | 08:42 |
| 3 | 100,100,100 | 0.9627 | 0.9191 | 0.0809 | 0.0373 | 0.9477 | 0.9627 | 0.9454 | 0.9551 | 0.98 | 07:51 |
| 4 | 300,300,300,300 | 0.9671 | 0.9137 | 0.0863 | 0.0329 | 0.9446 | 0.9671 | 0.9459 | 0.9557 | 0.979939 | 36:36 |
| 2 | 400,400 | 0.9755 | 0.901 | 0.099 | 0.0245 | 0.9375 | 0.9755 | 0.9459 | 0.9561 | 0.980394 | 28:08 |
| 4 | 200,200,200,200 | 0.9711 | 0.9107 | 0.0893 | 0.0289 | 0.9431 | 0.9711 | 0.9472 | 0.9569 | 0.981067 | 19:10 |
| 2 | 500,500 | 0.9693 | 0.9155 | 0.0845 | 0.0307 | 0.9459 | 0.9693 | 0.948 | 0.9575 | 0.980055 | 34:25 |
| 3 | 100,50,100 | 0.9707 | 0.9148 | 0.0852 | 0.0293 | 0.9455 | 0.9707 | 0.9486 | 0.958 | 0.980691 | 07:16 |
| 3 | 200,100,200 | 0.9658 | 0.9261 | 0.0739 | 0.0342 | 0.9521 | 0.9658 | 0.9501 | 0.9589 | 0.981751 | 15:04 |
| 3 | 100,200,100 | 0.9668 | 0.9251 | 0.0749 | 0.0332 | 0.9516 | 0.9668 | 0.9503 | 0.9591 | 0.982598 | 09:29 |
| 3 | 300,300,300 | 0.9649 | 0.9301 | 0.0699 | 0.0351 | 0.9546 | 0.9649 | 0.9511 | 0.9597 | 0.981693 | 25:08 |
| 3 | 500,500,500 | 0.9732 | 0.9248 | 0.0752 | 0.0268 | 0.9517 | 0.9732 | 0.954 | 0.9623 | 0.984339 | 48:29 |
| **3** | **400,400,400** | **0.9719** | **0.9272** | **0.0728** | **0.0281** | **0.9531** | **0.9719** | **0.9542** | **0.9624** | **0.983771** | **38:19** |

**Table 10**
Results for DL and seven machine learning algorithms (with stateful input generation and dynamic features only).

| | TPR | TNR | FPR | FNR | Precision | Recall | w-FM |
|---|---|---|---|---|---|---|---|
| NB | 0.62 | 0.855 | 0.145 | 0.38 | 0.765 | 0.768 | 0.764 |
| SL | 0.761 | 0.933 | 0.067 | 0.239 | 0.87 | 0.87 | 0.868 |
| SVM Linear | 0.758 | 0.938 | 0.062 | 0.242 | 0.872 | 0.872 | 0.87 |
| SVM RBF | 0.758 | 0.944 | 0.056 | 0.242 | 0.876 | 0.875 | 0.873 |
| J48 | 0.855 | 0.954 | 0.046 | 0.145 | 0.917 | 0.918 | 0.917 |
| PART | 0.861 | 0.955 | 0.045 | 0.139 | 0.92 | 0.92 | 0.92 |
| RF | 0.88 | 0.971 | 0.029 | 0.12 | 0.938 | 0.938 | 0.937 |
| DL(200,200,200) | 0.9776 | 0.9086 | 0.0914 | 0.0224 | 0.9482 | 0.9776 | 0.9627 |

achieved high accuracy performance reaching better figures than those presented in existing deep learning-based Android malware detection frameworks. To the best of our knowledge, this is the first work to investigate deep learning using dynamic features extracted from apps using real phones. Our results also highlight the significance of enhancing input generation for dynamic analysis systems that are designed to detect Android malware using machine learning. As future work, self-adaptation such as introduced and investigated recently for Intrusion Detection systems (Papamartzivanos et al., 2019) could be explored as a means of improving the performance of the deep learning based system for Android malware detection.

**Table 11**
Results for DL and seven machine learning algorithms (with stateful input generation and static + dynamic features).

|         | TPR    | TNR   | FPR   | FNR    | Precision | Recall | w-FM   |
|---------|--------|-------|-------|--------|-----------|--------|--------|
| NB      | 0.816  | 0.886 | 0.114 | 0.184  | 0.86      | 0.86   | 0.86   |
| SL      | 0.871  | 0.957 | 0.043 | 0.129  | 0.925     | 0.925  | 0.924  |
| SVM RBF | 0.871  | 0.957 | 0.043 | 0.129  | 0.925     | 0.925  | 0.924  |
| SVM Linear | 0.875 | 0.964 | 0.036 | 0.125 | 0.931   | 0.931  | 0.931  |
| J48     | 0.919  | 0.967 | 0.033 | 0.081  | 0.949     | 0.949  | 0.949  |
| PART    | 0.931  | 0.968 | 0.032 | 0.069  | 0.954     | 0.954  | 0.954  |
| RF      | 0.941  | 0.988 | 0.012 | 0.059  | 0.971     | 0.971  | 0.971  |
| DL(200,200,200) | 0.9956 | 0.967 | 0.033 | 0.0044 | 0.9809 | 0.9956 | 0.9882 |

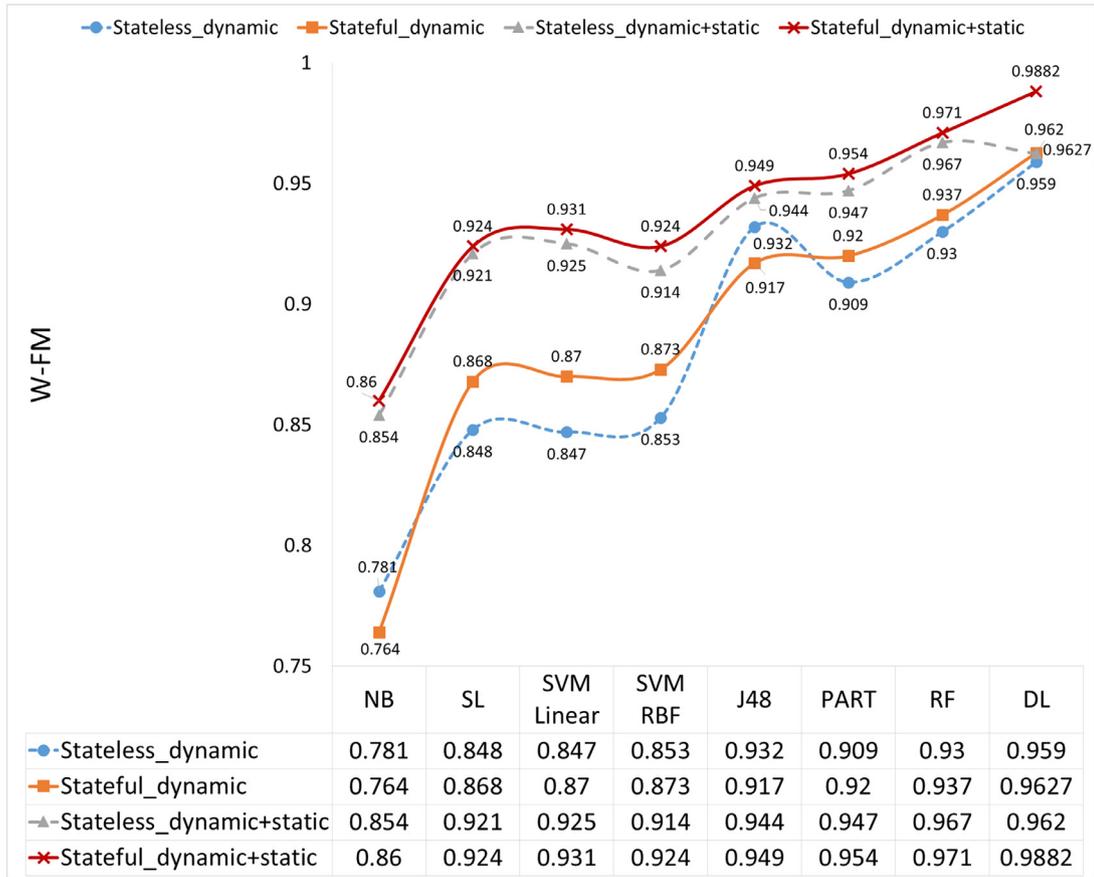

**Fig. 2.** w-FM for DL-Droid and seven selected ML algorithms. Stateful vs. stateless input generation.

**Table 12**
Comparisons of DL-Droid with other existing deep learning approaches.

| Classification System | Extracted Features Types | Benign | Malware | No. of neurons | Acc. | Prec. | Recall | F-score |
|---|---|---|---|---|---|---|---|---|
| DroidDetector Yuan et al. (2016) | Static only | 880 | 880 | [150,150] | 89.03 | 90.39 | 89.04 | 89.76 |
| DroidDetector Yuan et al. (2016) | Dynamic only | 880 | 880 | [150,150] | 71.25 | 72.59 | 71.25 | 71.92 |
| DroidDetector Yuan et al. (2016) | Static & Dynamic | 880 | 880 | [150,150] | 96.76 | 96.78 | 96.76 | 96.76 |
| CNN McLaughlin et al. (2017) | Static (opcode) | 863 | 1260 | N/A | 98 | 99 | 95 | 97 |
| MalDozer Karbab et al. (2017) | Static only | 37,627 | 20,089 | N/A | N/A | 96.29 | 96.29 | 96.29 |
| Deep4MalDroid Hou et al. (2016) | Dynamic (sys. Calls) | 1,500 | 1,500 | [200,200,200] | 93.68 | 93.96 | 93.36 | 93.68 |
| AutoDroid Hou et al. (2017) | Static only | 2,500 | 2,500 | [200,200,200] | 96.66 | 96.55 | 96.76 | 96.66 |
| DL-Droid (Stateless) | Dynamic only | 19,620 | 11,505 | [300,100,300] | 94.95 | 94.08 | 97.78 | 95.89 |
| DL-Droid (Stateless) | Static & Dynamic | 19,620 | 11,505 | [400,400,400] (Table 9) | 95.42 | 95.31 | 97.19 | 96.24 |
| DL-Droid (Stateful) | Dynamic only | 19,620 | 11,505 | [200,200,200] | 95.21 | 94.82 | 97.76 | 96.27 |
| DL-Droid (Stateful) | Static & Dynamic | 19,620 | 11,505 | [200,200,200] | 98.5 | 98.09 | 99.56 | 98.82 |

## Declaration of competing interest

The authors declare that they have no known competing financial interests or personal relationships that could have appeared to influence the work reported in this paper.

**Mohammed K. Alzaylaee** received his Ph.D. degree in Computer Science (Cyber Security) from Queen's University Belfast, U.K. in 2019. He Holds M.Sc. degree in Computer Science from the University of New Brunswick, Fredericton, NB, Canada, 2012, and currently he is an assistant professor in the Department of Information Systems, College of Computing in Al-Qunfudah, Umm-Al-Qura University, Saudi Arabia. His research interests include dynamic analysis of Android applications, machine learning for the malware detection, and cyber security.

**Suleiman Y. Yerima** received his Ph.D. degree in Mobile Computing and Communications in 2009 from the University of South Wales, U.K. (formerly University of Glamorgan). He holds an MSc (with Distinction) in Personal, Mobile and Satellite Communications from the University of Bradford, U.K and a B.Eng. (First Class) degree in Electrical and Computer Engineering from Federal University of Minna, Nigeria. He was a member of the Mobile Computing Communications and Networking (MoCoNet) Research group at Glamorgan from 2005 to 2009. From 2010 to 2012 he was a Research Assistant in the UK-India Advanced Technology Centre of excellence in Next Generation Networks, Systems and Services (IUATC) at the University of Ulster, Northern Ireland. He joined the Centre for Secure Information Technologies (CSIT), Queen's University Belfast in 2012 where he had been a Research Fellow in Mobile Security until 2017. He is currently a Senior Lecturer in Cyber Security at De Montfort University, Leicester, UK. Suleiman is a Certified Information Systems Security Professional (CISSP) and a Certified Ethical Hacker (CEH). He current research interests are in mobile security, malware analysis and detection, machine learning, intrusion detection systems and authentication. Suleiman is the recipient of the 2017 IET Information Security premium (best paper) award.

**Sakir Sezer** received the Dipl. Ing. degree in electrical and electronic engineering from RWTH Aachen University, Germany, and the Ph.D. degree in 1999 from Queen's University Belfast, U.K. Prof. Sezer is currently Secure Digital Systems Research Director and Head of Network Security Research in the School of Electronics Electrical Engineering and Computer Science at Queen's University Belfast. His research is leading major (patented) advances in the field of high performance content processing and is currently commercialized by Titan IC Systems. He has co-authored over 120 conference and journal papers in the area of highperformance network, content processing, and System on Chip. Prof. Sezer has been awarded a number of prestigious awards including InvestNI, Enterprise Ireland and Intertrade Ireland innovation and enterprise awards, and the InvestNI Enterprise Fellowship. He is also co-founder and CTO of Titan IC Systems and a member of the IEEE International System-on-Chip Conference executive committee.